\documentclass[%
 aip,
 amsmath,amssymb,
 reprint,%
]{revtex4-2}

\usepackage{xcolor}
\usepackage{graphicx}% Include figure files
\usepackage{dcolumn}% Align table columns on decimal point
\usepackage{bm}% bold math
\usepackage{hyperref}
\usepackage[utf8]{inputenc}
\usepackage[T1]{fontenc}
\usepackage{mathptmx}

\begin{document}

\newcommand{\hly}[1]{\colorbox{yellow}{\parbox{\columnwidth}{#1}}}

\preprint{AIP/123-QED}

\title[Electrical switching of antiferromagnetic CoO across the Néel temperature]{Electrical switching of antiferromagnetic CoO | Pt across the Néel temperature}
% Force line breaks with \\

\author{M. J. Grzybowski}
 \email{Michal.Grzybowski@fuw.edu.pl}
  \affiliation{Department of Applied Physics, Eindhoven University of Technology, 5600 MB Eindhoven, the Netherlands}
  \affiliation{Institute of Experimental Physics, Faculty of Physics, University of Warsaw, ul. Pasteura 5, PL-02-093 Warsaw, Poland}
 %\altaffiliation[Currently at: ]{Eindhoven University of Technology, 5600 MB Eindhoven, the Netherlands}
 %\affiliation{Eindhoven University of Technology, 5600 MB Eindhoven, the Netherlands}
\author{C. F. Schippers}%
 \affiliation{Department of Applied Physics, Eindhoven University of Technology, 5600 MB Eindhoven, the Netherlands}
\author{M. E. Bal}
\author{K. Rubi}
 \thanks{Current address: National High Magnetic Field Laboratory, Los Alamos National Laboratory, Los Alamos, NM 87545, USA}
\author{U. Zeitler}
 \affiliation{High Field Magnet Laboratory (HFML -EMFL), Radboud University, 6525 ED Nijmegen, the Netherlands}
\author{M. Foltyn}
  \affiliation{Institute of Physics, Polish Academy of Sciences, Aleja Lotnikow 32/46, Warsaw PL 02668, Poland}
\author{B.~Koopmans}
\author{H.~J.~M.~Swagten}
 \affiliation{Department of Applied Physics, Eindhoven University of Technology, 5600 MB Eindhoven, the Netherlands}

 %\homepage{http://www.Second.institution.edu/~Charlie.Author.}
%\affiliation{%
%School of Physics and Astronomy, University of Nottingham, United Kingdom%\\This line break forced% with \\
%}%

\date{\today}% It is always \today, today,

\begin{abstract}
One of the most important challenges in antiferromagnetic spintronics is the read-out of the Néel vector state. High current densities up to 10$^8$ Acm$^{-2}$ used in the electrical switching experiments cause notorious difficulty in distinguishing between magnetic and thermal origins of the electrical signals. To overcome this problem, we present a temperature dependence study of the transverse resistance changes in the switching experiment with CoO | Pt devices. We demonstrate the possibility to extract a pattern of spin Hall magnetoresistance for current pulses density of $5 \times 10^{7} \text{A}\,\text{cm}^{-2}$ that is present only below the Néel temperature and does not follow a trend expected for thermal effects. This is the compelling evidence for the magnetic origin of the signal, which is observed using purely electrical techniques. We confirm these findings by complementary experiments in an external magnetic field. Such an approach can allow determining the optimal conditions for switching antiferromagnets and be very valuable when no imaging techniques can be applied to verify the origin of the electrical signal.
\end{abstract}

\maketitle

Intensive interest in electrical manipulation of antiferromagnetism is caused by their potential appealing applications such as efficient and magnetic-field-robust data storage \cite{Olejnik2017NatComm} or long-range spin transport \cite{Lebrun2018Nature}. The electrical current can be used to control the spin axis of antiferromagnets. Numerous physical mechanisms can be responsible for that process \cite{Wadley2016Science,  Baldrati2019PRL, Moriyama2018SR, Chen2018PRL}. Experimental studies of antiferromagnetic oxides demonstrate the dominating role of the thermomagnetoelastic effect \cite{Zhang2019PRL, Meer2021NL, Baldrati2020PRL}, in which magnetic reorientation is driven by thermally induced strain. In principle, the reorientation of the N\'eel vector ($\bf{n}$) can be probed by either anisotropic magnetoresistance (AMR) or spin Hall magnetoresistance (SMR), the effects relying on resistivity dependence on the angle between $\bf{n}$ and current density or spin accumulation, respectively \cite{Hoogeboom2017APL, Fischer2018PRB, Baldrati2018PRB}. However, a notorious difficulty in electrical probing is the presence of non-magnetic thermal effects such as inhomogeneous heating or electromigration that can contribute to the measured resistance \cite{Chiang2019, Churikova2020, MatallaWagner2020}. It creates the need to use imaging techniques to provide the unambiguous proof of magnetic reorientation and show the correspondence between the electrical signals and the domains reorientation \cite{Grzybowski2017PRL, Wadley2018NN, Gray2019PRX, Meer2021NL}. Unfortunately, magnetic imaging techniques often exhibit complexity and surface sensitivity which limits their practical application in antiferromagnetic devices.

However, any electrical signal pattern of magnetic origin, that is resulting from the reorientation of antiferromagnetic spins, should be present only in magnetically ordered antiferromagnetic phase unlike non-magnetic thermal effects, that should be present in the ordered as well as in the disordered, paramagnetic state. Hence, in this study we propose to compare the electrical switching experiment at different temperatures, in particular below and above the N\'eel temperature $T_{\text{N}}$, by which we will gather crucial and unambiguous information about the presence of a genuine reorientation of the antiferromagnetic spin system. The antiferromagnetic CoO thin-films we are using have an experimentally easily accessible $T_{\text{N}}$ that is close to room temperature and exhibits in-plane biaxial anisotropy \cite{Cao2011APL, Baldrati2020PRL} making it a perfect material choice for the proposed study. Indeed, we determine that the magnetic reorientation in CoO | Pt can be detected electrically for current pulses of around $50\,\text{MA}\,\text{cm}^{-2}$ as in this regime the electrical patterns disappear above $T_{\text{N}}$. This is further confirmed with the electrical switching experiment in the external magnetic field. The results are critically important for complex structures comprising antiferromagnets (AFs), where the AF is only one of many layers buried far from the top interface \cite{Wang2015AM, Goto2016JJAP} and where the use of magnetic imaging is very challenging.

\begin{figure}
\includegraphics[width=0.65\linewidth]{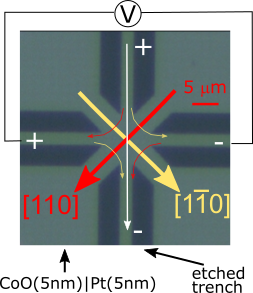}% Here is how to import EPS art
\caption{\label{fig:device}Micrograph of the device used for electrical switching experiments. The dark regions correspond to the insulating etched trenches whereas the remaining part is formed from a CoO layer grown on MgO (001) and capped with Pt. Red and yellow arrows symbolize the direction of the current pulses applied using 6 contacts, along magnetic easy axes of CoO $[110]$ and $[1\overline{1}0]$, respectively. The white arrow depicts the probing current direction.}
\end{figure}

To realize the experimental goal, 8 arm cross devices (Fig.~\ref{fig:device}) were prepared by e-beam lithography and ion milling from CoO thin-films of 5 nm thickness. They were grown by reactive DC magnetron sputtering at 430 $^\circ$C on MgO (001) single crystals with Pt top layer of 5 nm thickness that has been deposited at room temperature. The LEED pattern proves a good crystalline quality and XPS shows a spectrum typical for CoO \cite{supplement}. We adopt the value of $T_{\text{N}} = 272\pm{5}\,\text{K}$ that was determined by the point in which the spin-flop transition disappears in the temperature-dependent spin Hall magnetoresistance studies of analogous CoO | Pt samples \cite{supplement}. Therefore, an experiment performed in the temperature range $T \in [240\,\text{K}, 305\,\text{K}]$ will cover both interesting regimes: below and above the $T_\text{N}$ as well as in the vicinity of the critical point.

The electrical switching experiment consists of two parts. During the pulsing, a sequence of high-density current pulses $\mathbf{J}$ is applied. Each pulse has a length of $3\,\text{ms}$. At T = $240\,\text{K}$, each sequence consists of 5 pulses of $3\,\text{ms}$, separated by 5 seconds; at higher temperatures each sequence consists of a single pulse. The pulses are expected to cause a reorientation of the Néel vector $\bf{n}$. The pulses are applied along two orthogonal axes, $[110]$ or $[1\overline{1}0]$ (Fig.~\ref{fig:device}), which are the magnetic easy directions \cite{Cao2011APL, Baldrati2020PRL, supplement} that are preferred by spins. During the probing, transverse resistance is measured with a small current magnitude (0.25$\,\text{mA}$, corresponding to current density of $0.6\,\text{MAcm}^{-2}$) 1 second after each pulsing sequence using a lock-in amplifier. The reorientation of the N\'eel vector should be reflected in transverse resistance variations due to SMR. The experiment is performed for different magnitudes of current pulses $J$ and at different temperatures. The results are presented in the form of relative transverse resistivity changes $\Delta R_{\text{xy}}/R$ as a function of pulse number where $\Delta R_{\text{xy}}$ is the change of the transverse resistance upon reorientation of the pulsing direction and $R$ is mean longitudinal resistance which is $21.6\,\Omega$ at $T=240\,\text{K}$. The direction of the current pulse is changed between the two perpendicular easy axes after every 5 pulses whereas probing is always performed in the same configuration. The values of $J$ describing different experimental data sets are calculated by dividing the current that is stabilized during the pulsing by the narrowest cross-section.

\begin{figure}
\includegraphics[width=0.9\linewidth]{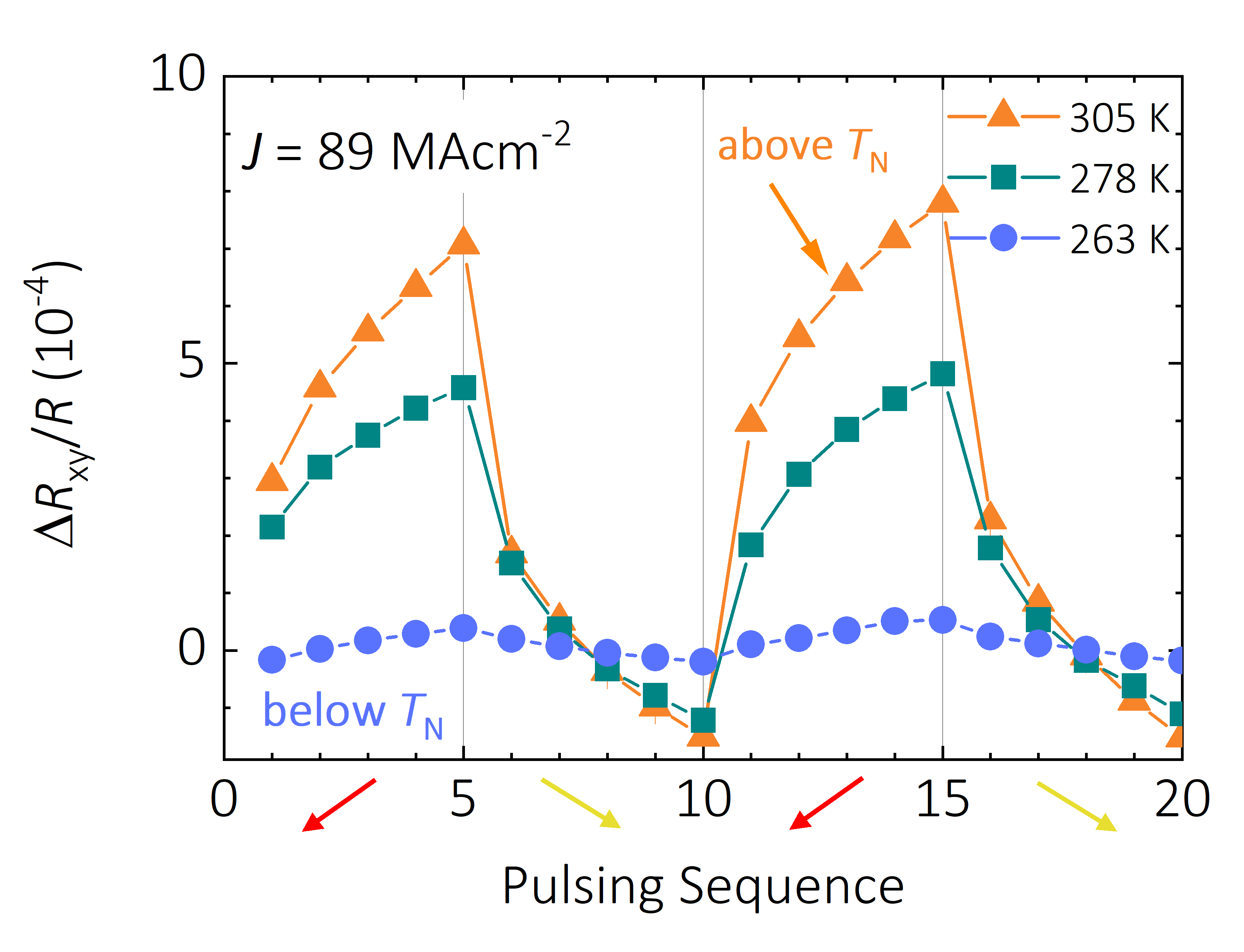}% Here is how to import EPS art
\caption{\label{fig:thermaleffects} Transverse resistance patterns recorded in electrical switching experiment for pulses with the highest current densities. The direction of the current pulse changes every 5 sequence as depicted by the red and yellow arrows that reflect the geometrical configuration presented in Fig.~\ref{fig:device}. The lines on the plot are guides for the eye.}
\end{figure} 
The experimental results for the largest current density in the range of $J_\text{t}\in [89, 92] \,\text{MA}\,\text{cm}^{-2}$ yield saw-tooth like patterns of transverse resistivity typical for thermal non-magnetic effects (Fig.~\ref{fig:thermaleffects}). This can be concluded from its presence above $T_{\text{N}}$ as shown by the data set collected at $T=305\,\text{K}$ (orange). Moreover, this type of signal exhibits a magnitude that increases with ambient temperature and exceeds $10^{-4}$. This magnitude is larger above $T_{\text{N}}$ than for the antiferromagnetic phase. No saturation of the signal with the increasing number of pulses of the same type was observed. Such behavior is caused by non-homogeneous current distribution within the device. The pulsing procedure creates non-equivalent current density points in the corners of the device \cite{Chiang2019}. It results in different local heating and non-zero transverse voltage during the probing. Apart from such temperature gradient \cite{Chiang2019, Churikova2020} also Seebeck effect \cite{Chiang2019} and structural changes like electromigration \cite{Churikova2020} or recrystallization \cite{MatallaWagner2020} may play an important role in contributing to the electrical probing signal.  For the smallest current magnitudes of pulsing (below $40\, \text{MA}\,\text{cm}^{-2}$) no significant variations of $\Delta R_{\text{xy}}/R$ is recorded at any temperature.

\begin{figure}
\includegraphics[width=0.9\linewidth]{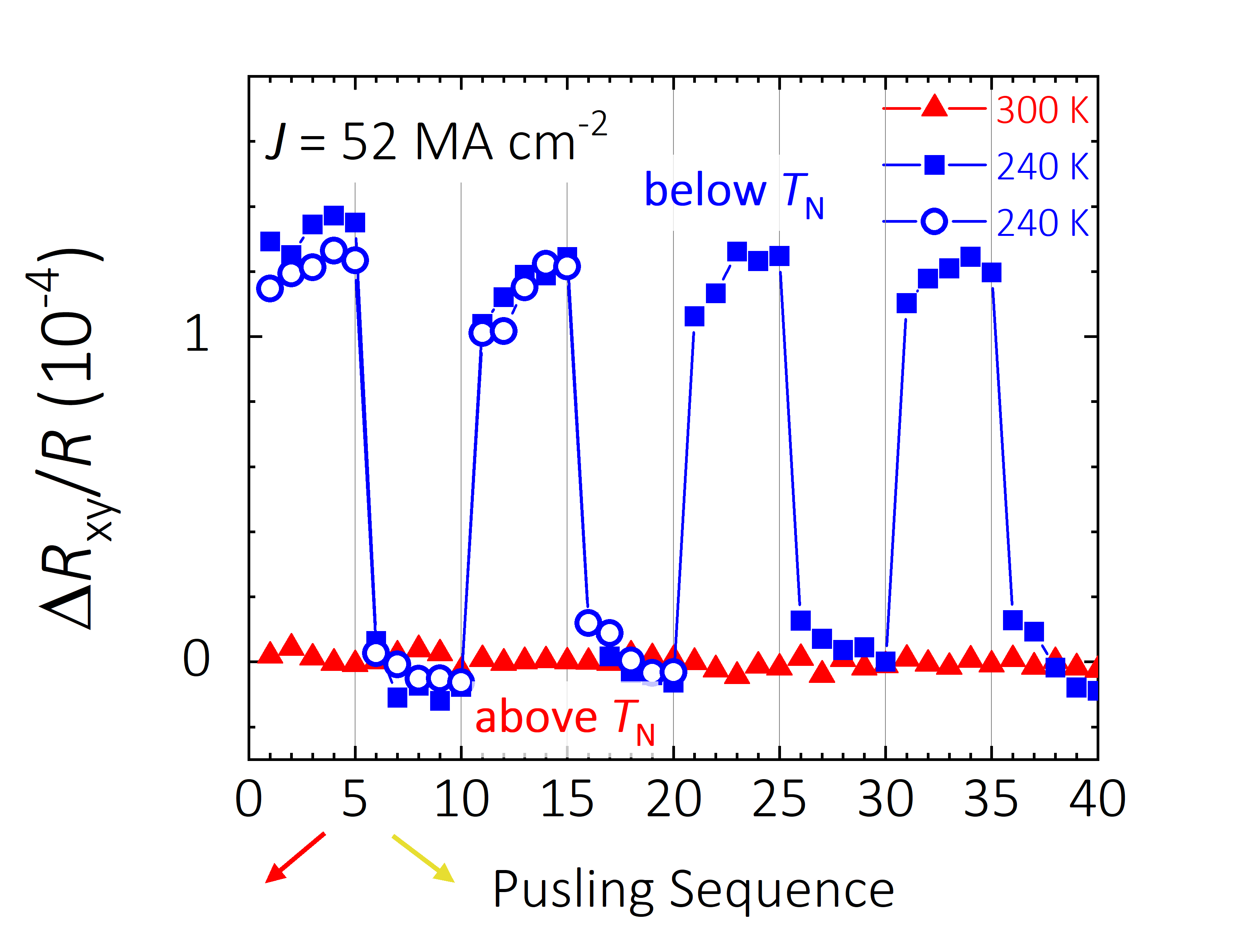}% Here is how to import EPS art
\caption{\label{fig:magneticeffect} Transverse resistance pattern recorded for $52 \,\text{MAcm}^{-2}$. A switching pattern is only present below $T_{\text{N}}$ (blue squares). The pattern is reproducible (white-blue circles) after several days of other switching experiments performed at higher $T$ and with larger $J$. The pattern is absent above $T_{\text{N}}$ (red). The lines on the plot are guides for the eye.}
\end{figure}

The most interesting results are obtained for intermediate values of the current pulse magnitude. Close to $J_\text{m} \in [47,56]\,\text{MA}\,\text{cm}^{-2}$ a pattern of $\Delta R_{\text{xy}}/R$ appears, which is present only below $T_{\text{N}}$ (Fig.~\ref{fig:magneticeffect}). The pattern collected for this range of current magnitudes is reproducible after days of performing multiple experiments with higher currents and at higher temperatures. Its presence below $T_{\text{N}}$ and absence above it can be a strong indication of its magnetic origin in contrast to the data recorded for the highest current densities (Fig~\ref{fig:thermaleffects}).

\begin{figure}
\includegraphics[width=1\linewidth]{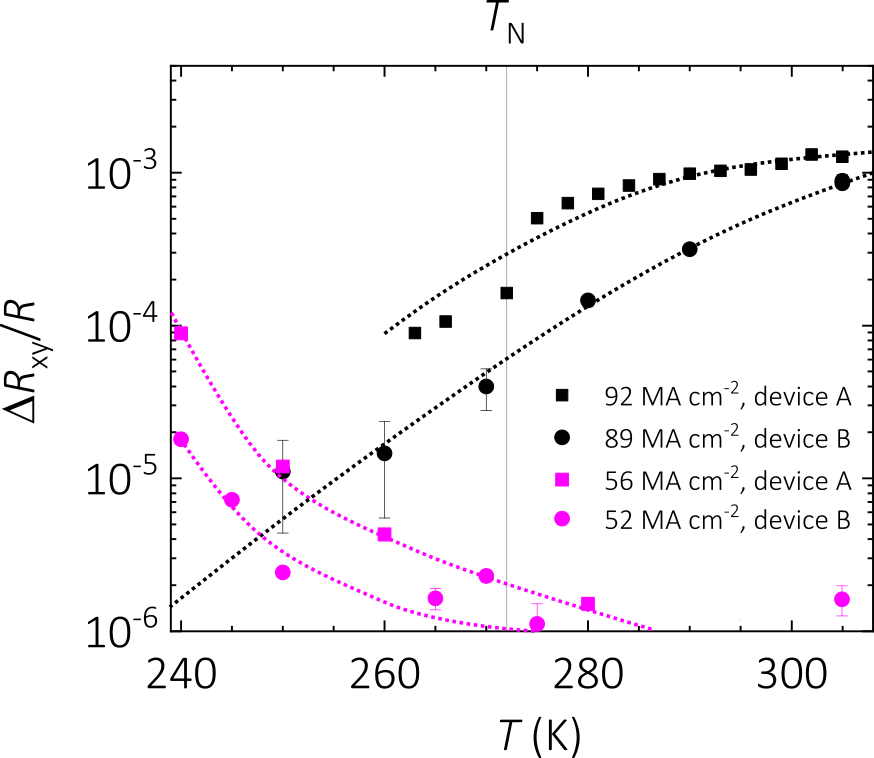}% Here is how to import EPS art
\caption{\label{fig:tempdep} Relative transverse resistance changes $\Delta R_{\text{xy}}/R$ in electrical switching experiment as a function of temperature recorded for $J_\text{m}$ (pink) and $J_\text{t}$ (black) and two different devices A and B (squares and circles, respectively). The lines are guides for the eye.}
\end{figure}

To highlight that distinct physical mechanisms are responsible for the appearance of the signal in different current density regimes, the temperature dependence of the transverse resistance patterns in the electrical switching experiment are summarized in Fig.~\ref{fig:tempdep}. It shows that the thermal effects, crucial in the highest current densities ($J_\text{t}$), result in the increasing magnitude of the signal with temperature. This is not the case for the moderate current densities ($J_\text{m}$), where the signal is clearly decreasing between $T\in[240\,\text{K},\,260\,\text{K}]$ and it is practically undetectable above $T_{\text{N}}$. These tendencies are reproducible among different devices (Fig.~\ref{fig:tempdep}). They allow suspecting the resistance variations detected for $J_\text{m}$ are caused by the spin reorientation in AF CoO. This is finally verified in the switching experiment in the external magnetic field.

Unambiguous proof of the magnetic origin of the electrical signal for $J_\text{m}$ is provided by the switching experiment in the external magnetic field. The reorientation of spins in an antiferromagnet can be suppressed by a strong magnetic field that sets the AF in the spin flop state. It can occur for $B$ strong enough to overcome the anisotropy.  For a thin-films of CoO at $T=240\,\text{K}$ the spin flop field state can be reached at $7\,\text{T}$ when $B\parallel [110]$ \cite{supplement}. Even if electrical current switches the spin axis of the antiferromagnet it should immediately come back to the initial state due to the spin flop induced by the magnetic field once the current pulse is turned off. Meanwhile, thermal effects, in principle, are not dependent on the magnetic field. Therefore, by studying the magnetic field dependence of the electrical switching patterns it is possible to identify the magnetic component of the signal. Similar approaches to use a magnetic field to validate the origin of the electrical signal have been used in hematite \cite{Zhang2019PRL} and CoO \cite{Baldrati2020PRL}.

\begin{figure}
\includegraphics[width=0.9\linewidth]{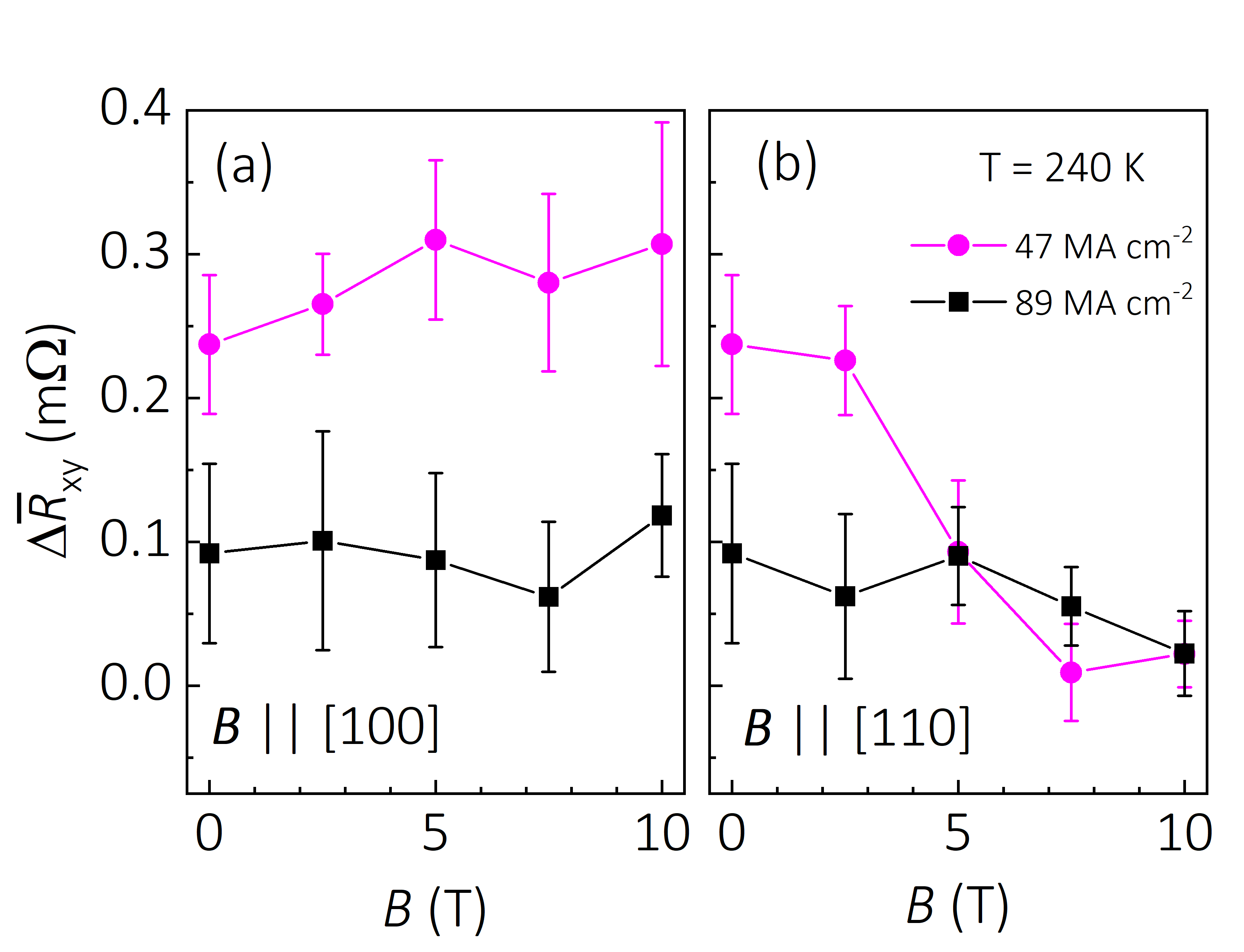}% Here is how to import EPS art
\caption{\label{fig:infield} The magnetic field dependence of the mean transverse resistance changes $\Delta \overline{R_{\text{xy}}}$ in the switching experiment recorded for $50 \,\text{MA}\,\text{cm}^{-2}$ (violet) and $90 \,\text{MA}\,\text{cm}^{-2}$ (black) for the magnetic field parallel to the (a) $[100]$ hard magnetic direction and (b) $[110]$ easy magnetic direction. The experiment was conducted at $T=240\,\text{K}$. Lines on the plot are guides for the eye.}
\end{figure}

The in-field experiment shows that the electrical switching pattern for $J_\text{m}$ disappears upon increasing magnetic field (Fig.~\ref{fig:infield}b), whereas the thermal pattern remains nearly unchanged as expected. The resistance changes for $J_\text{m}$ at $5\,\text{T}$ are diminished and above this value they become undetectable. Moreover, the response of the system to the magnetic field is anisotropic - such dependence is not observed for $B \parallel [100]$ (Fig.~\ref{fig:infield}a). It is in agreement with the anisotropy of thin films of CoO, in which $[100]$ is a magnetic hard direction and the magnetic field of $10\,\text{T}$ does not cause perpendicular alignment $\mathbf{n} \perp \mathbf{B}$ in this geometrical configuration \cite{supplement}.

Therefore, the signal obtained for $J_\text{m}$ can be interpreted as spin Hall magnetoresistance (SMR). The sign of the resistance variations corresponds to setting $\mathbf{n} \parallel \mathbf{J}$. This is different from another study of CoO electrical switching \cite{Baldrati2020PRL} and may suggest that an antidamping torque mechanism is responsible for the antiferromagnetic reorientation \cite{Chen2018PRL}. However, it has to be noted that even the same thermomagnetoelastic mechanism can result in different end states of $\mathbf{n}$ depending on device design and experimental geometry \cite{Meer2021NL}.

Finally, we estimate the temperature increase of the CoO layer in the device region due to the electrical pulses by a 3D finite element method model \cite{sup}. It reveals that the $T_N$ is not exceeded during the pulsing at $T=240\,\text{K}$ presented in Fig.~\ref{fig:magneticeffect} and the temperature rise for $50\,\text{MAcm}^{-2}$ is around $7\,\text{K}$. It means that the device remains in the antiferromagnetic state during the whole experiment. It is also true even for such high current densities as $92\,\text{MAcm}^{-2}$ at $T=240\,\text{K}$. Most of the heat is dissipated in the first few milliseconds after the pulsed current is turned off. Still, a slight temperature increase is present even for long timescales typical for probing. Together with the symmetry of the device design, this can produce patterns of the electrical signal visible at elevated current densities and at high temperatures as shown in Fig.~\ref{fig:thermaleffects}.

In summary, we demonstrate that it is possible to determine whether a signal in an electrical switching experiment of an antiferromagnet is of magnetic origin, using purely electrical techniques. The method relies on comparing the signal at temperatures above and below the Néel temperature. Finally, we validated the study by performing the experiment in a strong external magnetic field. Such an approach can allow determining the optimal experimental conditions for switching antiferromagnets and be very valuable when no imaging techniques can be applied.

We acknowledge J. Francke, B. van der Looij and G. Basselmans for technical help; R. Lavrijsen and M. Zgirski for helpful discussions and J. Dedding for the measurements of $T_N$. We acknowledge support from HFML-RU, member of the European Magnetic Field Laboratory (EMFL). Sample fabrication was performed using NanoLabNL facilities. The research was funded by the Dutch Research Council (NWO) under Grant 680-91-113. The research was funded in part by National Science Centre, Poland under Grant 2021/40/C/ST3/00168 and grant 2019/34/E/ST3/00432. For the purpose of Open Access, the author has applied a CC-BY public copyright licence to any Author Accepted Manuscript (AAM) version arising from this submission.
\nocite{*}
\bibliography{TempSwitchingCoOv4}% Produces the bibliography via BibTeX.

\end{document}